\documentclass[a4paper,11pt]{article}

\usepackage{jcappub} 

\usepackage[T1]{fontenc} 
\usepackage[utf8]{inputenc}
\usepackage{siunitx}
\usepackage{todonotes}

\title{\boldmath Estimation of an Upper Limit on the Density of Relic Neutrinos in the Sun via the Solar $^8$B Neutrino Flux}

\author[a,1]{T. Ruhe,\note{Corresponding author.}}
\author[a,2]{A. Sandrock\note{Now at National Research Nuclear University MEPhI, Moscow, Russia.}}

\affiliation[a]{Technische Universität Dortmund,\\Dortmund, Germany}

\emailAdd{tim.ruhe@udo.edu}
\emailAdd{alexander.sandrock@udo.edu}

\abstract{Relic neutrinos from the early universe are predicted to have a relatively large number density, but extremely low energies. Hence, the only possible interaction proceeds via neutrino capture on beta-decaying nuclei. In case relic neutrinos are captured by beta-decaying nuclei in the Sun, the neutrinos normally emerging from the decay of these nuclei will be missing from the overall number of solar neutrinos registered in neutrino experiments. Within the Sun, ${}^8\mathrm{B}$ and three nuclei from the CNO cycle are found to be suitable for this kind of process. Their cross section as well as the possible impact on the observed number of solar neutrinos are discussed. Assuming standard neutrino oscillations, no deviations of neutrino flux measurements from the predictions of the solar standard model are observed and upper limits are derived accordingly. }

\newcommand{\Boron}{$^{8}$B}
\newcommand{\CnuB}{C$\nu$B}
\newcommand{\Nitrogen}{$^{13}$N}
\newcommand{\Oxygen}{$^{15}$O}
\newcommand{\Fluor}{$^{17}$F}

\begin{document}

\maketitle
\flushbottom

\section{Introduction}

Similar to the well established Cosmic Microwave Background (CMB), a background of relic neutrinos with a temperature of approximately 1.9\,K -- the so-called C$\nu$B -- is one of the expected consequences of the Big Bang. Although the number density of these neutrinos is expected to be on the order of $50$ per $\si{\centi\meter\cubed}$, their experimental detection is far from trivial and has not been achieved yet. Challenges in the detection of the C$\nu$B mainly arise from the small energies of the neutrinos, which are in the \si{\micro eV} range. These small energies directly correspond to extremely small interaction cross sections ($\sigma_{NCB}v_{\nu} < \SI{1e-41}{cm^2}$). Furthermore, in many cases the energy of the neutrinos is not sufficient to reach the ground state of a neighbouring nucleus in an inverse beta decay. 

However, a possible detection mechanism, originally proposed by Weinberg~\cite{Weinberg1962}, has recently gained some momentum due to the PTOLEMY project~\cite{Ptolemy2013}. In this detection channel, neutrinos interact with beta-decaying nuclei, thereby eliminating the requirement of carrying enough energy to reach the ground- or any excited state in the neighbouring nucleus. Instead, this neutrino capture on beta-decaying nuclei can proceed unhindered. In addition this reaction offers a unique experimental signature consisting of a delta peak located at electron energies of $E=Q_{\beta}+2m_{\nu}$, where $Q_{\beta}$ corresponds to the $Q$-value of the beta-decay and $m_{\nu}$ represents the mass of the neutrino. Due the sub-eV mass of the neutrino, a direct detection of this process remains a highly challenging endeavour. 

 Although the solar neutrino flux originates  from various processes within the pp-chain, solar neutrinos detected in experiments such as Super-Kamiokande~\cite{SuperK2008}, SNO~\cite{SNO2001} and Borexino~\cite{Borexino2010} solely originate from the $\beta^+$-decay of \Boron, an experimental limitation arising from the energy threshold of the individual experiments. The fluxes registered at the aforementioned experiments are consistent with theoretical expectations, considering the Standard Solar Model (SSM) and standard neutrino oscillations, within their respective uncertainties. In this paper we show that this agreement and the experimental uncertainties can be utilized to derive upper limits on the number density of relic neutrinos captured in the Sun. 

The paper is organized as follows: Section~\ref{sec:status} provides the necessary theoretical background and briefly reviews the experimental results on the solar neutrino flux. In Sec.~\ref{sec:limits}, the upper limits on C$\nu$B neutrinos in the Sun are derived and possible improvements possible via measurements of neutrinos from the solar CNO cycle are discussed. Section~\ref{sec:discussion} concludes the paper with a discussion of the results.

\section{Theoretical and Experimental Considerations}
\label{sec:status}

A possible detection mechanism for relic neutrinos, originally proposed by Weinberg, is the neutrino capture on beta-decaying nuclei (NCB), which proceeds via the following equation~\cite{Weinberg62}:
\begin{equation}
    \nu_e + ^{A}_{Z}X \longrightarrow ^{A}_{Z+1}Y + e^{-}.    
	\label{eq:ncb}
\end{equation}
The actual reaction rates, however, depend on the properties of the nucleus, for example on whether the decay is super-allowed, allowed or forbidden~\cite{Cocco2007}. Cocco et al. have studied the properties of approximately 1500 beta-decaying nuclei with respect to their use in NCB-experiments~\cite{Cocco2007}. Not unexpected, allowed and super-allowed beta-decays yield the highest cross sections for NCB. For these nuclei the ratio of NCB- to the beta-decay-rate is given by
\begin{equation}
    \dfrac{\lambda_{\nu}}{\lambda_{\beta}} = \left ( \lim_{p_{\nu \to 0}} \sigma_\text{NCB}v_{\nu} \right) n_{\nu} \dfrac{t_{1/2}}{\ln 2},
\label{eq:capture_ratio}
\end{equation}
where $n_{\nu}$ represents the number density of neutrinos and $t_{1/2}$ is the half-life of the beta-decaying nucleus. $\lambda_{\nu}$ and $\lambda_{\beta}$ represent the beta-decay and neutrino capture rate, respectively. The neutrino interaction cross section times neutrino velocity 
is denoted as $ \sigma_{\text{NCB}}v_{\nu}$. 

Solar neutrinos are produced in abundance in various beta-decays in the pp-chain, but only neutrinos from the $\beta^+$-decay of \Boron\,have been experimentally detected in solar neutrino experiments. \Boron\,decays with a half-life of $t_{1/2}=\SI{0.77}{\second}$ to $^{8}$Be. The $Q$-value with respect to the ground state of $^{8}$Be is $Q=\SI{17979}{\kilo\eV}$. It should, however, be noted that the decay into the ground-state of $^{8}$Be is second order forbidden ($2^+\rightarrow 0^+$). The decay proceeds predominantly ($\approx100$\%~\cite{TOI}) via the first excited state located at an excitation energy of $E_{X,1}=\SI{3040}{\kilo\eV}$. In this case, the considered decay is allowed ($2^+ \rightarrow 2^+$) and the effective $Q$-value is altered to  $Q_{\textrm{eff}}=\SI{14939}{\kilo\eV}$.

Although the CNO-cycle is expected to play a subdominant role and neutrinos from beta-decaying nuclei within this cycle have not yet been detected, the properties of the involved nuclei (\Nitrogen, \Oxygen\,and \Fluor) with respect to NCB are worth discussing. All three nuclei decay via allowed $\beta^{+}$-decay into the ground states of their daughter-nuclei. Furthermore, $Q$- and $\log(ft)$-values are also comparable. Compared to \Oxygen\,and \Fluor, which have half-lifes of $\SI{122.24}{\second}$ and $\SI{64.29}{\second}$, \Nitrogen\,is found to have a significantly larger half-life of $t_{1/2}=\SI{9.965}{\minute}$. For all three nuclei the $\beta^+$-decay is an allowed transition ($1/2^- \rightarrow 1/2^-$ for \Nitrogen~and \Oxygen, $5/2^+\rightarrow 5/2^+$ for \Fluor). The properties of \Boron,\Nitrogen, \Oxygen\,and \Fluor\, are summarized in Tab.~\ref{tab:nuclei_infos}. 
\begin{table}[t]
\begin{center}
\begin{tabular}[t]{| l | l | l | l | l |}
\hline 
Nucleus & $Q$-value in keV & $t_{1/2}$ & $\log(ft)$ & $\sigma_\text{NCB}v_{\nu}/c$ in cm$^2$\\
\hline 
$^{8}$B & \num{14939} & 0.77 s & 3.3 & \num{1.32e-41}  \\
$^{13}$N & \num{2220.4} & 9.965 min & 3.66 & \num{1.85e-43} \\
$^{15}$O & \num{2753.9} & 122.24 s & 3.64 & \num{2.84e-43} \\
$^{17}$F & \num{2760.7} & 64.29 s & 3.36 & \num{5.56e-43} \\
\hline 
\end{tabular}

\caption{$Q$-values, half-lifes, $\log(ft)$ values and cross section for $\beta^{+}$-decaying nuclei in the Sun.}
\label{tab:nuclei_infos}

\end{center}
\end{table}

In this paper we consider eight different results on the flux of solar neutrinos and their respective statistical and systematic uncertainties. The results taken into account were obtained with Super-Kamiokande I to IV~\cite{SuperK2006,SuperK2008,SuperK2011,SuperK2016}), Borexino~\cite{Borexino2010,Borexino2017} and SNO~\cite{SNO2001,SNO2004,SNO2008,SNOCombined}. As expected, statistical and systematic uncertainties were found to decrease with increasing measurement time and refined analyses. The most sensitive measurements with respect to obtaining limits on the relic neutrinos in the Sun are a combined analysis of all four phases of Super-Kamiokande~\cite{SuperK2016} and a recent result from Borexino~\cite{Borexino2017}, which report fluxes of $2.55^{+0.17}_{-0.19} (\textrm{stat.}) \pm0.07 (\textrm{syst.}) \times \SI{e6}{\per\centi\meter\squared\per\second}$ and $2.345\pm0.014(\textrm{stat.}) \pm 0.036 (\textrm{syst.})\times \SI{e6}{\per\centi\meter\squared\per\second}$, respectively.

As all four of the investigated nuclei undergo $\beta^{+}$-decay, the $\nu_e$ in Eq.~\eqref{eq:ncb} needs to be replaced with a $\bar{\nu}_e$ and an $e^{+}$ rather than a $e^-$ is observed in the final state. The limits derived in the next section, therefore only account for the $\bar{\nu}_e$-component of the \CnuB. 

\section{Upper Limits on the C$\nu$B neutrinos in the Sun}
\label{sec:limits}

In order to estimate an upper limit on relic neutrinos in the Sun,  Eq.~\ref{eq:capture_ratio} has to be solved for $n_{\nu}$, which yields:
\begin{equation}
    n_{\nu}=\dfrac{\lambda_{\nu}}{\lambda_{\beta}}\dfrac{\ln2}{t_{1/2}\left ( \lim_{p_{\nu \to 0}}  \sigma_{NCB}v_{\nu} \right)}.
\label{eq:density}    
\end{equation}
The density of relic neutrinos in the Sun is thus proportional the ratio of $\beta$-decay and NCB rates. This ratio, however, is not known experimentally, but one can conclude that it is smaller than the uncertainties reported in solar neutrino measurements, as the fluxes obtained in these measurements were found to be consistent with the Solar Standard Model, within those uncertainties, assuming standard neutrino oscillations. Furthermore, $n_{\nu}$ is found to be proportional to the inverse of $\sigma_{NCB}v_{\nu} t_{1/2}$, and therefore expected to vary from nucleus to nucleus. For the remainder of this section, upper limits obtained for \Boron-neutrinos from the pp-chain and limits possibly achievable via the detection of CNO-neutrinos will be discussed separately. The utilized cross sections were obtained using the recipe provided in~\cite{Cocco2007} and are summarized in Table~\ref{tab:nuclei_infos}.

\subsection{\Boron-neutrinos}
\label{sec:boron}
\begin{figure}[ttt]
\centering 
\includegraphics[width=.9\textwidth]{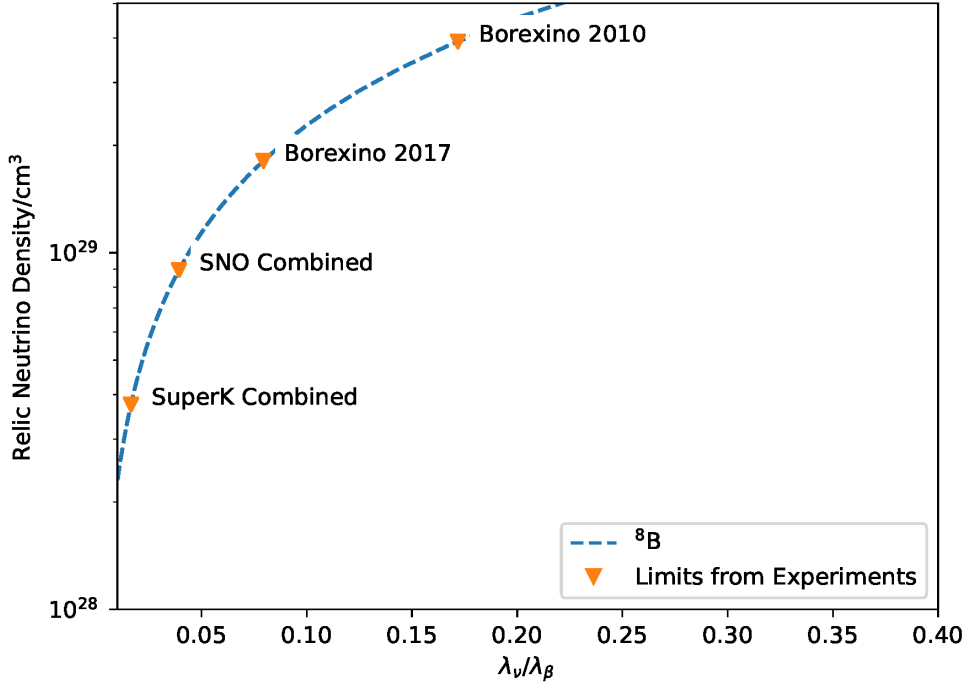}
\caption{\label{fig:limits_all} Limits obtained considering the uncertainties reported by Super-Kamiokande~\cite{SuperK2016}, SNO~\cite{SNOCombined} and Borexino~\cite{Borexino2010,Borexino2017}.}
\end{figure}
The results reported in this section, were ontained using the uncertainties reported by the respective experiments, which are summarized in Tab.~\ref{tab:limits} together with the obtained limits. To obtain the results, statistical and systematic uncertainties were added in quadrature. For cases where the upper and the lower bound on the uncertainty were found to be different, the larger value was used. 

Figure~\ref{fig:limits_all} shows the limits obtained using uncertainties reported by Super-Kamiokande~\cite{SuperK2016}, SNO~\cite{SNOCombined} and Borexino~\cite{Borexino2010,Borexino2017}. The obtained limits are depicted by the orange data points, whereas the blue dashed line represents the limits obtainable for \Boron-neutrinos as a function of the experimental uncertainty.

One finds that the best limit of $n_{\nu}\leq 3.75\times10^{28}\,\si{\per\cm\cubed}$ was obtained for the uncertainties reported in a combined measurement, using all four phases of the Super-Kamiokande experiment. Using the uncertainties of a combined analysis of the SNO data yields an upper limit of $n_{\nu}\leq 8.94\times10^{28}\,\si{\per\cm\cubed}$. Limits of $n_{\nu}\leq 18.1\times10^{28}\,\si{\per\cm\cubed}$ and $n_{\nu}\leq 39.1\times10^{28}\,\si{\per\cm\cubed}$ were obtained using the uncertainties reported for Borexino in 2010 and 2017, respectively.
\begin{figure}[ttt]
\centering 
\includegraphics[width=.9\textwidth]{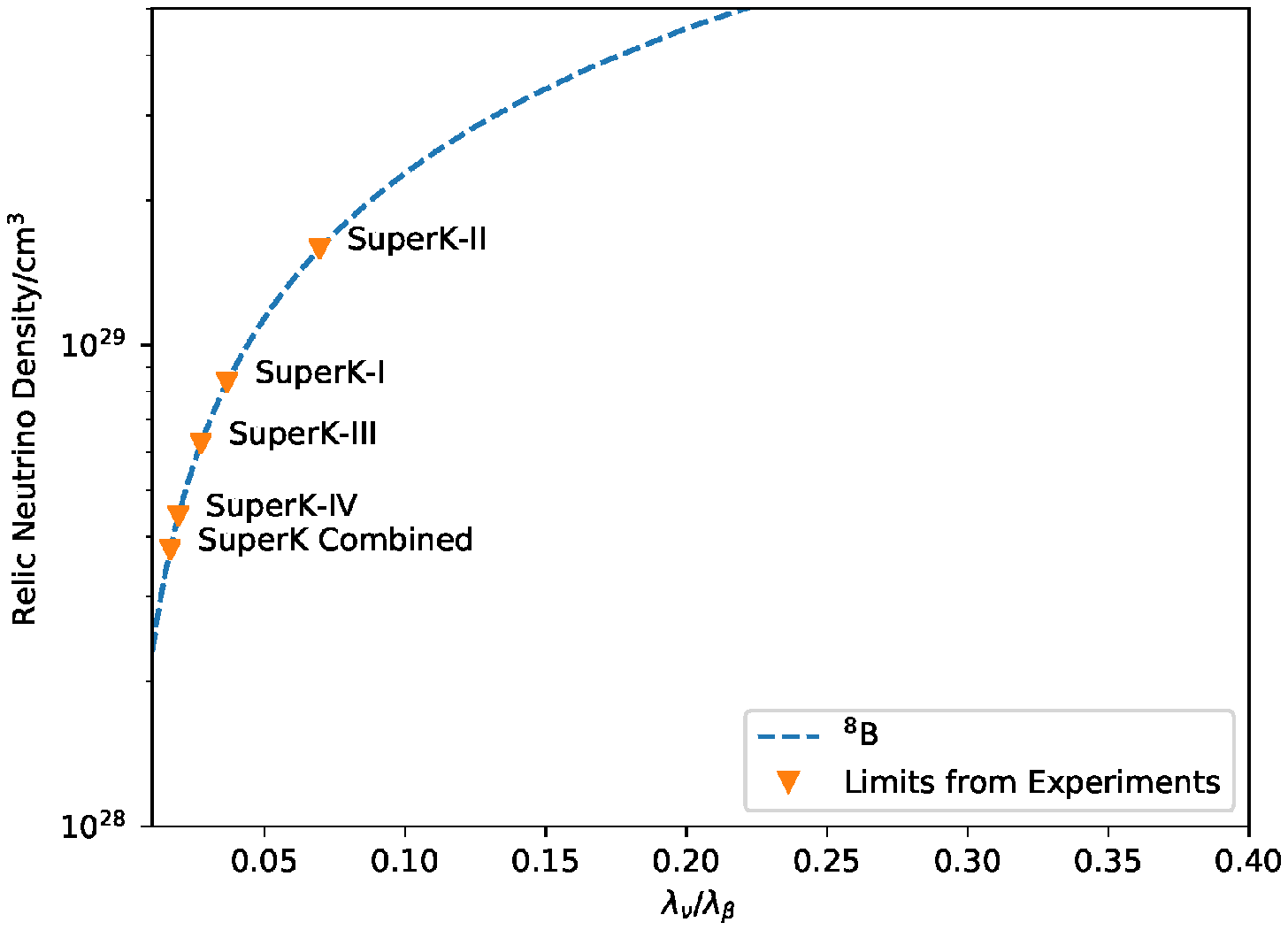}
\caption{\label{fig:limits_SuperK} Limits obtained for the four phases of Super-Kamiokande and an analysis using data from all four phases (labeled SuperK Combined). }
\end{figure}

Figure~\ref{fig:limits_SuperK} shows the evolution of the uncertainties for different phases of the Super-Kamiokande experiment. As expected the uncertainties and in turn also the limits on \CnuB-neutrinos decrease with time. Considering the uncertainties of the phase IV results of Super-Kamiokande, a limit of $n_{\nu}\leq 4.41\times10^{28}\,\si{\per\cm\cubed}$ is obtained. This limit is only marginally larger than the one obtained for the combined analysis. Somewhat larger limits are obtained, when considering the results from phase I and III ($n_{\nu}\leq 8.36\times10^{28}\,\si{\per\cm\cubed}$ for phase I and $n_{\nu}\leq 6.23\times10^{28}\,\si{\per\cm\cubed}$ for phase III). Using the results from phase II yields the largest limit ($n_{\nu}\leq 15.8\times10^{28}\,\si{\per\cm\cubed}$) for Super-Kamiokande, as uncertainties of $\pm 0.05 (\textrm{stat.}) ^{+0.16}_{-0.15} (\textrm{syst.})\times10^{6}\,\si{\per\centi\meter\squared\per\second}$ were reported. The differences in the uncertainties between Super-Kamiokande I and II arise from the different lifetimes of the samples (791 days for SuperKamiokande II and 1496 days for SuperKamiokande I) and differences in energy scale, event selection and reconstruction methods~\cite{SuperK2008}. 

\begin{figure}[ttt]
\centering 
\includegraphics[width=.9\textwidth]{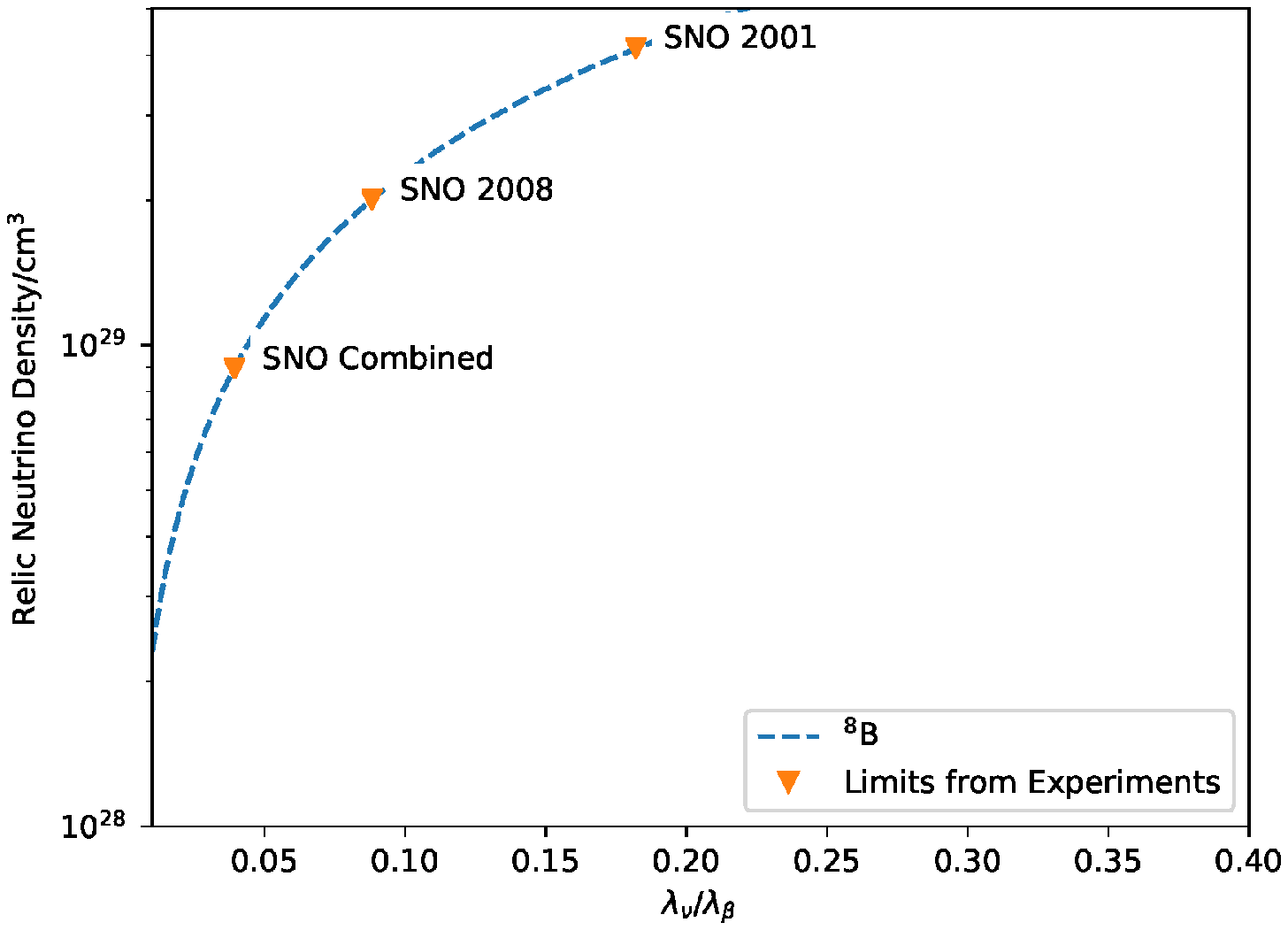}
\caption{\label{fig:limits_SNO} Limits on the density of \CnuB-neutrinos in the Sun, obtained for different experimental uncertainties, reported by the SNO-collaboration.}
\end{figure}

Figure~\ref{fig:limits_SNO} shows the evolution of the uncertainties as well as their impact on the limit of relic neutrinos in the Sun for different analyses carried out with the SNO. As for Super-Kamiokande the uncertainties are found to decrease with time and the smallest statistical and systematic uncertainties are obtained for a combined analysis ($\pm 0.16 (\textrm{stat.}) ^{+0.11}_{-0.13} (\textrm{syst.}$~\cite{SNOCombined}). Using this uncertainty, one obtaines a limit of $n_{\nu}\leq 8.94\times10^{28}\,\si{\per\cm\cubed}$, which is approximately a factor of two larger than the one obtained for the combined analysis of the Super-Kamiokande data. 

The differences in uncertainty between the measurements reported in 2004 and 2008 are only marginal~\cite{SNO2004,SNO2008} (see Table~\ref{tab:limits} for detailed values), thus, only the results obtained for the 2008 measurement are shown in the plot, in order to increase the visibility. Accordingly, the obtained limits -- $n_{\nu}\leq 20.4\times10^{28}\,\si{\per\cm\cubed}$ for 2004 and $n_{\nu}\leq 20.1\times10^{28}\,\si{\per\cm\cubed}$ for 2008 are also similar. Using the uncertainties from~\cite{SNO2001} yields the worst limit ($n_{\nu}\leq 41.5\times10^{28}\,\si{\per\cm\cubed}$ ), due to the large uncertainty of $\pm 0.99\times10^{28}\,\si{\per\cm\squared\per\second}$. 

\begin{table}[t]

\begin{center}
\begin{tabular}[t]{| l | l | l |}
\hline 
Experiment & Uncertainty in $\times10^{6}\,\si{\per\centi\meter\squared\per\second}$ & Limit $\times 10^{28}\,\si{\per\cm\cubed}$ \\
\hline 
Borexino 2010 & $\pm 0.4 (\textrm{stat.}) \pm 0.1 (\textrm{syst.})$ & 39.1 \\
Borexino 2017 & $^{+0.17}_{-0.19} (\textrm{stat.}) \pm0.07 (\textrm{syst.})$ & 18.1 \\
SuperK Combined & $\pm 0.014 (\textrm{stat.}) \pm 0.1 (\textrm{syst.})$ & 3.75 \\
SuperK-I & $\pm 0.02 (\textrm{stat.}) \pm 0.036 (\textrm{syst.})$ & 8.36 \\
SuperK-II & $\pm 0.05 (\textrm{stat.}) ^{+0.16}_{-0.15} (\textrm{syst.})$ &  15.8 \\
SuperK-III & $\pm 0.04 (\textrm{stat.}) \pm 0.05 (\textrm{syst.})$ & 6.23 \\
SuperK-IV & $\pm 0.02 (\textrm{stat.}) ^{+0.0.039}_{-0.040} (\textrm{syst.})$ & 4.41 \\
SNO 2001 & $\pm 0.99$ & 41.5 \\
SNO 2004 & $\pm 0.27 (\textrm{stat.}) \pm 0.38 (\textrm{syst.})$ & 20.4 \\
SNO 2008 & $^{+0.33}_{-0.31} (\textrm{stat.}) ^{+0.36}_{-0.34} (\textrm{syst.})$ & 20.1 \\
SNO Combined & $\pm 0.16 (\textrm{stat.}) ^{+0.11}_{-0.13} (\textrm{syst.})$ & 8.94 \\
\hline 
\end{tabular}

\caption{Uncertainties and the respective limits on \CnuB-neutrinos for different measurements of the solar \Boron-neutrino flux.}
\label{tab:limits}
\end{center}
\end{table}

\subsection{CNO-neutrinos}
\label{sec:cno}
\begin{figure}[ttt]
\centering 
\includegraphics[width=.9\textwidth]{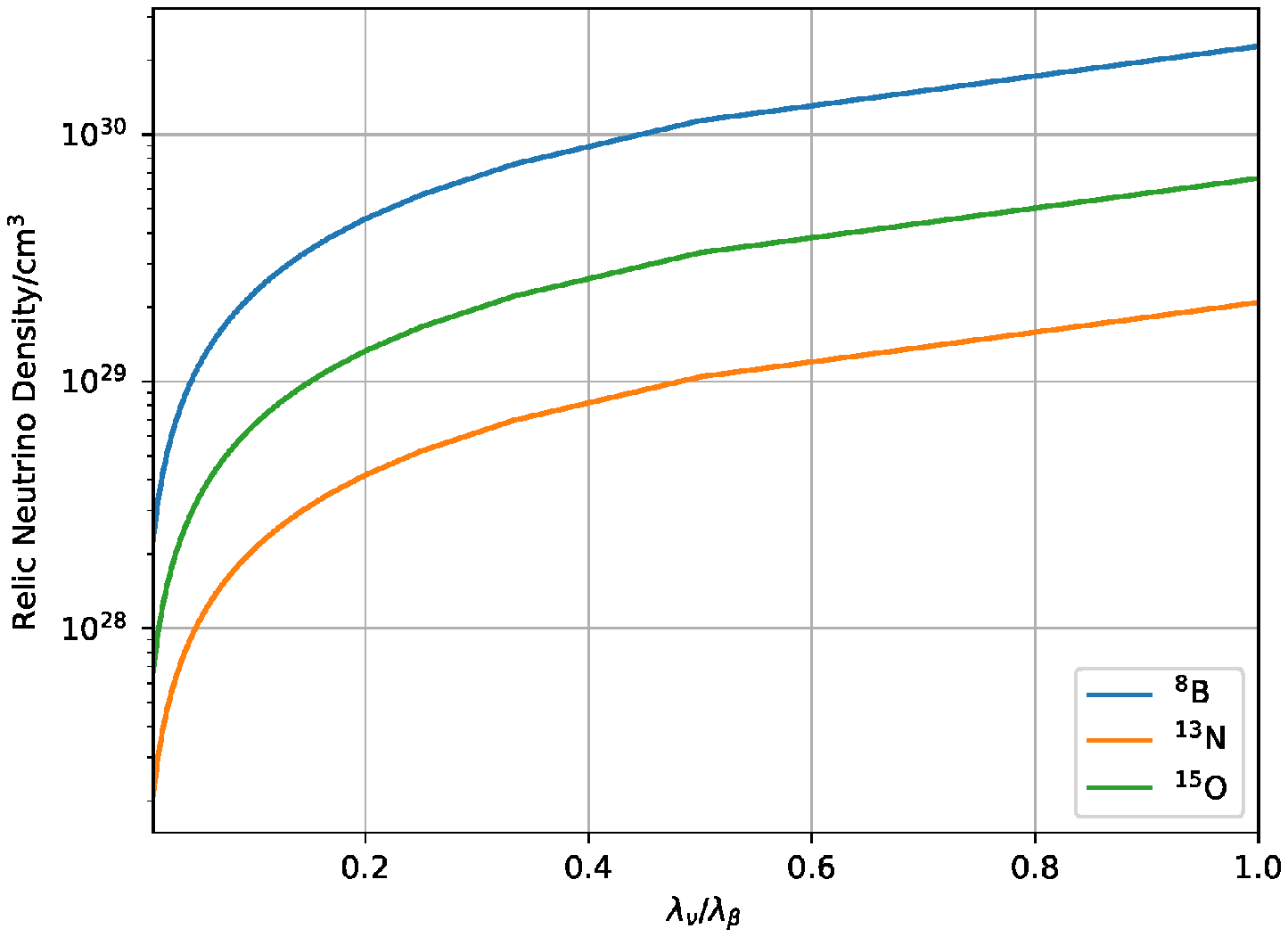}
\caption{\label{fig:limits_cno} Upper limits of the relic neutrino density in the Sun as a function of the measurement uncertainty.}
\end{figure}

Figure~\ref{fig:limits_cno} shows the obtainable limits for neutrinos from the CNO-cycle as a function of the experimental uncertainty. Although neutrino emission is expected from three beta-decaying nuclei within the CNO-cycle (\Nitrogen, \Oxygen, \Fluor), only \Nitrogen\, and \Oxygen\, are depicted. The achievable limits for \Oxygen\,are shown in green, whereas the possible limits for \Nitrogen\,are shown in orange. The blue line represents the upper limits obtained for \Boron\,and is shown for comparison. Values obtained for \Fluor\,are not depicted to increase the visibility in the plot. Limits for \Fluor, however, are very similar to the ones obtained for \Oxygen. The half life of \Oxygen\,is approximately a factor of two larger compared to \Fluor, which compensated by the NCB cross section, which is approximately a factor of two smaller. The $Q$-values of the two nuclei differ only marginally. As \Nitrogen~has the smallest NCB cross section of the four nuclei considered here, the small upper limit solely arises from the comparably large half life of $9.965$\,minutes, which is one order of magnitude larger than the half life of \Boron.

From Fig.~\ref{fig:limits_cno} one finds that the achievable limits on \CnuB-neutrinos in the Sun are about one order of magnitude smaller than the ones obtained from current solar neutrino measurements. The current best limit, obtained using uncertainties from a combined analysis of Super-Kamiokande data, for example, can be reached by measuring neutrinos from the decay of \Oxygen\,or \Fluor\,with a total relative uncertainty of $\lambda_{\nu} / \lambda_{\beta}=5.6$\%. With the detection of neutrinos from \Nitrogen\,a similar limit can be obtained with a relative uncertainty as large as $\lambda_{\nu} / \lambda_{\beta}=18$\%. 

\section{Summary and Discussion}
\label{sec:discussion}

In the preceding sections it was shown that a limit on the density of relic neutrinos in the Sun can be obtained, using the statistical and systematic uncertainties of solar neutrino measurements, assuming standard neutrino oscillations and the Standard Solar Model. The obtained results depend on the uncertainties reported by the individual experiments, and the best upper limit of $n_{\nu}\leq 3.75\times10^{28}\,\si{\per\cm\cubed}$ was derived using the uncertainties reported in a combined analysis of data taken in all four phases of the Super-Kamiokande experiment. Limits for all experimental results considered in this paper are summarized in Tab.~\ref{tab:limits}.

While limits on the density of relic neutrinos in the Sun can be derived from the agreement of the experimental results with the parameters allowed in the Solar Standard Model, the opposite, however, does not hold true. Possible differences in the observed fluxes from theoretical predictions are far more likely to be explained by differences in one or more parameters of the Solar Standard Model, e.g. its metallicity. 

Nevertheless, the obtained best limit can be used to check whether the result is reasonable. Using the best limit from this paper and taking the radius of the Sun as $r_{\texttt Sun} = 6.9551\times10^{10}\,\si{\centi\meter}$ we find that the maximum number of electron neutrinos from the \CnuB\,in the Sun is $N_{\nu}\leq 5.285\times10^{61}$. Further, assuming an upper limit on the mass of the electron neutrino of $m_{\nu} \leq 1.1\,\si{\eV}$~\cite{Katrin2019} one finds that the contribution of \CnuB-neutrinos to the mass of the Sun is smaller than $1.0362\times10^{26}\,\si{\kilo\gram}$, which corresponds to a relative contribution smaller than $5.21\times10^{-5}$. Considering that the neutrino is the lightest standard model particle, and that the Sun predominantly consists of Hydrogen and Helium, this is a reasonable result. 

Taking the ratio of $\nu_e:\nu_{\mu}:\nu_{\tau}$ in the \CnuB\,to $1:1:1$, the limits from Sec.~\ref{sec:limits} need to be scaled by a factor of 3 or 6, in order to obtain the total density of \CnuB-neutrinos, depending on, whether the neutrino is a Dirac- or Majorana particle. It was further shown that the achievable limits can be significantly improved by considering beta-decaying nuclei from the CNO-cycle. Compared to \Boron\,the obtainable limits are smaller by approximately a factor of 5 for \Oxygen\,and\Fluor. For \Nitrogen\,the limit would improve by one order of magnitude.

\bibliographystyle{plain}
\bibliography{references.bib}

\begin{thebibliography}{10}

\bibitem{SuperK2011}
K.~{Abe} et~al.
\newblock {Solar neutrino results in Super-Kamiokande-III}.
\newblock {\em \prd}, 83(5):052010, March 2011.

\bibitem{SuperK2016}
K.~{Abe} et~al.
\newblock {Solar neutrino measurements in Super-Kamiokande-IV}.
\newblock {\em \prd}, 94(5):052010, September 2016.

\bibitem{Borexino2017}
M~{Agostini} et~al.
\newblock {Improved measurement of 8B solar neutrinos with 1.5 kt y of Borexino
  exposure}.
\newblock {\em arXiv e-prints}, September 2017.

\bibitem{SNO2008}
B.~{Aharmim} et~al.
\newblock {Independent Measurement of the Total Active B8 Solar Neutrino Flux
  Using an Array of He3 Proportional Counters at the Sudbury Neutrino
  Observatory}.
\newblock {\em Physical Review Letters}, 101(11):111301, September 2008.

\bibitem{SNOCombined}
B.~{Aharmim} et~al.
\newblock {Combined analysis of all three phases of solar neutrino data from
  the Sudbury Neutrino Observatory}.
\newblock {\em \prc}, 88(2):025501, August 2013.

\bibitem{SNO2001}
Q.~R. {Ahmad} et~al.
\newblock {Measurement of the Rate of {$\nu$}$_{e}$ + d $\rightarrow$ p + p +
  e$^{-}$ Interactions Produced by $^{8}$B Solar Neutrinos at the Sudbury
  Neutrino Observatory}.
\newblock {\em Physical Review Letters}, 87(7):071301, August 2001.

\bibitem{SNO2004}
S.~N. {Ahmed} et~al.
\newblock {Measurement of the Total Active $^{8}$B Solar Neutrino Flux at the
  Sudbury Neutrino Observatory with Enhanced Neutral Current Sensitivity}.
\newblock {\em Physical Review Letters}, 92(18):181301, May 2004.

\bibitem{Katrin2019}
M~Aker et~al.
\newblock An improved upper limit on the neutrino mass from a direct kinematic
  method by katrin.
\newblock {\em arXiv preprint arXiv:1909.06048}, 2019.

\bibitem{Borexino2010}
G.~{Bellini} et~al.
\newblock {Measurement of the solar B8 neutrino rate with a liquid scintillator
  target and 3 MeV energy threshold in the Borexino detector}.
\newblock {\em \prd}, 82(3):033006, August 2010.

\bibitem{Ptolemy2013}
S.~{Betts} et~al.
\newblock {Development of a Relic Neutrino Detection Experiment at PTOLEMY:
  Princeton Tritium Observatory for Light, Early-Universe, Massive-Neutrino
  Yield}.
\newblock {\em arXiv e-prints}, page arXiv:1307.4738, Jul 2013.

\bibitem{Cocco2007}
A.~G. Cocco, G.~Mangano, and M.~Messina.
\newblock Probing low energy neutrino backgrounds with neutrino capture on beta
  decaying nuclei.
\newblock {\em Journal of Physics: Conference Series}, 110(8):082014, may 2008.

\bibitem{SuperK2008}
J.~P. {Cravens} et~al.
\newblock {Solar neutrino measurements in Super-Kamiokande-II}.
\newblock {\em \prd}, 78(3):032002, August 2008.

\bibitem{TOI}
Richard Firestone.
\newblock {\em Table of Isotopes, 8th Edition}.
\newblock Wiley-VCH), 1999.

\bibitem{SuperK2006}
J.~{Hosaka} et~al.
\newblock {Solar neutrino measurements in Super-Kamiokande-I}.
\newblock {\em \prd}, 73(11):112001, June 2006.

\bibitem{Weinberg62}
S.~Weinberg.
\newblock The neutrino problem in cosmology.
\newblock {\em Il Nuovo Cimento}, 25(1):15--27, July 1962.

\bibitem{Weinberg1962}
Steven Weinberg.
\newblock Universal neutrino degeneracy.
\newblock {\em Phys. Rev.}, 128:1457--1473, Nov 1962.

\end{thebibliography}








\end{document}